\documentclass[11pt]{article}
\usepackage{amsmath,amsthm,amscd,array,amssymb}

\begin{document}

\begin{center}
{\Large{\bf 
$N_T = 8$, $D=2$ Hodge--type cohomological gauge theory
\\
\medskip\smallskip
with global $SU(4)$ symmetry}}\footnote{Talk given at
3. Int. Andrei Sakharov Conference on Physics, 
Moscow, June 24 - 29, 2002 }
\\
\bigskip\medskip
{\large{\sc B. Geyer}}$^a$
\footnote{Email: geyer@itp.uni-leipzig.de}
{\large {\sc and }}
{\large{\sc D. M\"ulsch}}$^{b}$
\footnote{Email: muelsch@informatik.uni-leipzig.de}
\\
\smallskip
{\it $^a$ Universit\"at Leipzig, Naturwissenschaftlich-Theoretisches Zentrum
\\
$~$ and Institut f\"ur Theoretische Physik, D--04109 Leipzig, Germany
\\\smallskip
$\!\!\!\!\!^b$ Wissenschaftszentrum Leipzig e.V., D--04103 Leipzig, Germany}
\\
\bigskip
{\small{\bf Abstract}}
\\
\end{center}

\begin{quotation}
\noindent {\small{We show that the partially topological twisted $N = 16$, 
$D = 2$ super Yang--Mills theory gives rise to  a $N_T = 8$ 
Hodge--type cohomological gauge theory with global $SU(4)$ symmetry.}}
\end{quotation}

\bigskip
%%%%%%%%%%%%%%%%%%%%%%%%%%%%%%%%%%%%%%%%%%%%%%%%%%%%%%%%%%%%%%%%%%%%%%%%%%%%%
\begin{flushleft}
{\large{\bf 1. Introduction}}
\end{flushleft}
%%%%%%%%%%%%%%%%%%%%%%%%%%%%%%%%%%%%%%%%%%%%%%%%%%%%%%%%%%%%%%%%%%%%%%%%%%%%%
\medskip

Some very enlightening, but preliminary attempts have been made to incorporate 
into the gauge--fixing procedure of general gauge theories besides the basic 
ingredience of BRST cohomology $\Omega$ also a co--BRST cohomology 
$^\star \Omega$ which, together with the BRST Laplacian $W$, form the same 
kind of superalgebra as the de Rham cohomology operators in differential 
geometry (for a review, see, e.g., Ref.~\cite{1}). This would allow, 
according to the Hodge--type decomposition 
$\psi = \omega + \Omega \chi + \,^\star \Omega \phi$
of a general quantum state,
by imposing both the BRST condition $\Omega \psi = 0$ and
the co--BRST condition $^\star \Omega \psi = 0$ upon $\psi$, 
to select the uniquely determined harmonic state $\omega$ thereby
projecting onto the subspace of physical states.

It has been a long--standing problem to present a non--abelian field 
theoretical model obeying such a Hodge--type cohomological structure. 
Recently, the authors have shown \cite{2} that the
dimensional reduced Blau--Thompson model \cite{3} ---
the novel $N_T = 2$ topological twist of the $N = 4$, $D = 3$ super Yang--Mills
theory (SYM) --- gives a prototype example of a $N_T = 4$, $D=2$ 
 Hodge--type cohomological 
gauge theory. The conjecture, that topological gauge theories 
could be possible candidates for Hodge--type cohomological 
theories was already asserted by van Holten \cite{4}. In fact,
 $D = 2$ topological gauge theories \cite{5} are of particular interest
 because of their relation to
$N = 2$ superconformal theories \cite{6} and  Calabi--Yau moduli 
spaces \cite{7}. 

Here we present another example of a Hodge--type cohomological 
gauge theory. It is obtained by a $N_T = 8$ topological twist of the Euclidean 
$N = 16$, $D = 2$ SYM, and its  action  localizes onto the moduli space of 
complexified flat connections.
The $N_T = 8$ scalar supercharges $Q^\alpha$ and $^\star Q^\alpha$ of 
that theory form a topological superalgebra which is completely analogous to 
the de Rham cohomology. Both supercharges are interrelated by a discrete 
Hodge--type $\star$ operation and generate the topological shift and co--shift 
symmetries. In accordance with the group theoretical description of some 
classes of topologically twisted low--dimensional supersymmetric world--volume 
theories \cite{3}, it is shown that this $N_T = 8$ cohomological theory has 
actually the global symmetry group $SU(4)$. Such effective low--energy 
world--volume theories appear quite naturally in the study of curved D--branes 
and D--brane instantons wrapping around supersymmetric cocycles for special 
Lagrangian submanifolds of Calabi--Yau $n$--folds (see, e.g., \cite{8,9,3}).

The paper is organized as follows: 
In Sec. 2 we briefly describe the BRST complex of general gauge theories 
based on harmonic gauges. 
In Sec. 3 we obtain the Euclidean $N = 16$, $D = 2$ SYM theory 
with R--symmetry group $SO(8)$ from the $N = 4$, $D = 4$ SYM via 
dimensional reduction to $D = 2$.
In Sec. 4 we perform the partiall $N_T=8$ topological twist of this SYM theory 
thereby getting the looked for $N_T = 8$ 
Hodge--type cohomological theory with global symmetry group $SU(4)$.  
 A more detailed  version will be presented elsewhere \cite{9a}.

\bigskip
%%%%%%%%%%%%%%%%%%%%%%%%%%%%%%%%%%%%%%%%%%%%%%%%%%%%%%%%%%%%%%%%%%%%%%%%%%%%%
\begin{flushleft}
{\large{\bf 2. BRST complex and Hodge decomposition}}
\end{flushleft}
%%%%%%%%%%%%%%%%%%%%%%%%%%%%%%%%%%%%%%%%%%%%%%%%%%%%%%%%%%%%%%%%%%%%%%%%%%%%%
\medskip

In order to select uniquely the physical states from the ghost--extended 
quantum state space some attempts \cite{1} have been made to 
incorporate into the gauge--fixing procedure of general gauge theories besides 
the BRST cohomology $\Omega$ also a co--BRST cohomology $^\star \Omega$ 
which, together with the BRST Laplacian $W$, obeys the following BRST--complex:
\begin{equation*}
\Omega^2 = 0, 
\qquad 
^\star \Omega^2 = 0,
\qquad
[ \Omega, W ] = 0, 
\qquad  
[ \,^\star\Omega, W ] = 0,
\qquad
W = \{ \Omega, \,^\star \Omega \} \neq 0,
\end{equation*}
where $\Omega$ and $^\star\Omega$ have ghost number $+1$ and $-1$,
respectively. Obviously, $^\star\Omega$ can not be identified with the 
anti--BRST operator $\bar\Omega$ which anticommutes with $\Omega$.

Representations of this algebra for the first time have been considered by 
Nishijima \cite{10}. However, since $\Omega$ and $^\star\Omega$ are 
nilpotent hermitian operators they cannot be realized in a Hilbert space. 
Instead, the BRST complex has to be represented in a Krein space $\cal K$ \cite{9b}.
$\cal K$ is obtained from a Hilbert space $\cal H$  with non--degenerate positive inner
product $(\chi , \psi)$ if $\cal H$ will be endowed also with a self--adjoint 
metric operator $J \neq 1$, $J^2 = 1$, allowing for the introduction of 
another non--degenerate, but indefinite scalar product 
$\langle \chi | \psi \rangle := ( \chi ,J \psi )$.
With respect to the inner product $\Omega$ and $^\star \Omega =
\pm J \Omega J$ are adjoint to each other, $( \chi , \,^\star \Omega \psi ) 
= ( \Omega \chi , \psi )$, however they are self--adjoint with respect to the
indefinite scalar product of $\cal K$. Notice, that different inner products 
$( \chi , \psi )$ lead to different co--BRST operators!

From these definitions one obtains a remarkable correspondence between the 
BRST cohomology and the de Rham cohomology:
\begin{alignat*}{4}
&\hbox{BRST operator}
&\quad&
\Omega, 
&\qquad\qquad&
\hbox{differential}
&\quad&
d,
\\
&\hbox{co--BRST operator}
&\quad&
^\star\Omega = \pm J \Omega J,
&\qquad\qquad&
\hbox{co--differential}
&\quad&
\delta = \pm \star d \star,
\\
&\hbox{duality operation}
&\quad&
J,
&\qquad\qquad&
\hbox{Hodge star}
&\quad&
\star,
\\
&\hbox{BRST Laplacian}
&\quad&
W = \{ \Omega, \,^\star\Omega \},
&\qquad\qquad&
\hbox{Laplacian}
&\quad& 
\Delta = \{ d, \delta \}.
\end{alignat*}
Because of this correspondence one denotes a state $\psi$
to be BRST (co--)closed iff $\Omega \psi = 0$ ($^\star\Omega \psi = 0$), 
BRST (co--)exact iff $\psi = \Omega \chi$ ($\psi = \,^\star\Omega \phi$) 
and BRST harmonic iff $W \psi = 0$. Completely analogous to the Hodge 
decomposition theorem in differential geometry there exists a corresponding
decomposition of any state  $\psi$ into a harmonic, an exact and a co--exact state,
$\psi = \omega + \Omega \chi + \,^\star \Omega \phi$.
The physical properties
of $\psi$ lie entirely within the BRST harmonic part $\omega$ which is given
by the zero modes of the operator $W$; thereby $W \omega = 0$ 
implies $\Omega \omega = 0 = \,^\star\Omega \omega$, and vice versa. The 
cohomologies of the (co--)BRST operator are given by equivalence classes:
\begin{alignat*}{2}
{\rm H}(\Omega) &= \frac{{\rm Ker}\, \Omega}{{\rm Im}\, \Omega}, 
&\qquad 
\psi \sim \psi' &= \psi + \Omega \chi \quad
(\hbox{equivalence class}),
\\
{\rm H}(\,^\star\Omega) &= \frac{{\rm Ker} \,^\star\Omega}
{{\rm Im} \,^\star\Omega}, 
&\qquad 
\psi \sim \psi' &= \psi + \,^\star\Omega \phi \quad 
(\hbox{equivalence class}).
\end{alignat*}
By imposing only the BRST gauge condition, $\Omega \psi = 0$,
within the equivalence class of BRST--closed states $\psi = \omega +
\Omega \chi$ besides the harmonic state $\omega$ there occur also 
spurious BRST--exact states, $\Omega \chi$, which have zero physical
norm. On the other hand, by imposing also the co--BRST gauge condition,
$^\star\Omega \psi = 0$, one gets for each BRST cohomology class the
uniquely determined harmonic state, $\psi = \omega$. 

\bigskip
%%%%%%%%%%%%%%%%%%%%%%%%%%%%%%%%%%%%%%%%%%%%%%%%%%%%%%%%%%%%%%%%%%%%%%%%%%%%%
\begin{flushleft}
{\large{\bf 3. Dimensional reduction of the $N=4$, $D=4$ super
Yang--Mills theory}}
\end{flushleft}
%%%%%%%%%%%%%%%%%%%%%%%%%%%%%%%%%%%%%%%%%%%%%%%%%%%%%%%%%%%%%%%%%%%%%%%%%%%%%
\medskip

Our final aim is to show that by a partial topological twist of $N = 16$, $D = 2$
SYM one gets a $N_T = 8$ Hodge--type cohomological theory with global symmetry 
group $SU(4)$. However, since the relationship between the twisted 
and untwisted fields is rather complex, let us first introduce the $N=16, D=2$ SYM.
This theory can be obtained by dimensional reduction to $D = 2$ from either
$N = 1$, $D = 10$ SYM or $N = 4$, $D = 4$ SYM. Because the latter
theory is well known, we choose the last possibility. 

The field content of $N = 4$, $D = 4$ 
SYM consists of an anti--hermitean gauge field $A_\mu$, two Majorana spinors
$\lambda_{A \alpha}$ and $\bar{\lambda}_{\dot{A}}^{\!~~\alpha}$ 
($\alpha = 1,2,3,4$) which transform as the fundamental and its complex 
conjugate representation of $SU(4)$, respectively, and a set of complex scalar 
fields $G_{\alpha\beta} = \hbox{$\frac{1}{2}$} 
\epsilon_{\alpha\beta\gamma\delta} G^{\gamma\delta}$, which transform as the 
second--rank complex selfdual representation of $SU(4)$. All the 
fields take their values in the Lie algebra $Lie({\cal G})$ of some compact 
gauge group $\cal G$.

In Euclidean space this theory has the following invariant action \cite{11}:
\begin{align}
\label{3.1}
S^{(N = 4)} = \int_E d^4x\, {\rm tr} \Bigr\{&
\hbox{$\frac{1}{4}$} F_{\mu\nu} F^{\mu\nu} -  
i \bar{\lambda}_{\dot{A}}^{\!~~\alpha} (\sigma_\mu)^{\dot{A} B} 
D^\mu \lambda_{B \alpha} + 
\hbox{$\frac{1}{64}$} [ G_{\alpha\beta}, G_{\gamma\delta} ]
[ G^{\alpha\beta}, G^{\gamma\delta} ] 
\nonumber
\\
& - \hbox{$\frac{1}{2}$} i \lambda_{A \alpha} 
[ G^{\alpha\beta}, \lambda^A_{\!~~\beta} ] -
\hbox{$\frac{1}{2}$} i \bar{\lambda}^{\dot{A} \alpha} 
[ G_{\alpha\beta}, \bar{\lambda}_{\dot{A}}^{\!~~\beta} ] +
\hbox{$\frac{1}{8}$} D_\mu G_{\alpha\beta} D^\mu G^{\alpha\beta} \Bigr\},
\end{align}
where the numerically invariant tensors $(\sigma_\mu)^{A \dot{B}}$ and 
$(\sigma_\mu)_{\dot{A} B}$ are the Clebsch--Cordon coefficients relating 
the representation $(1/2,1/2$ of $SL(2,C)$ to the
the vector representation of $SO(4)$,,
\begin{alignat}{2}
\label{3.2}
&(\sigma_\mu)^{\dot{A} B} = ( -i \sigma_1, -i \sigma_2, -i \sigma_3, I_2 ),
&\qquad 
&(\sigma_\mu)_{\dot{A} B} \equiv (\sigma_\mu)^{\dot{C} D}
\epsilon_{\dot{C}\dot{A}} \epsilon_{DB} = (\sigma_\mu^*)^{\dot{A} B},
\nonumber
\\
&(\sigma_\mu)_{A \dot{B}} = ( i \sigma_1, i \sigma_2, i \sigma_3, I_2 ),
&\qquad
&(\sigma_\mu)^{A \dot{B}} \equiv \epsilon^{AC} \epsilon^{\dot{B}\dot{D}}
(\sigma_\mu)_{C \dot{D}} = (\sigma_\mu^*)_{A \dot{B}},
\end{alignat}
$(\sigma_\mu)_{\dot{A} B}$ and $(\sigma_\mu)^{A \dot{B}}$ being the
corresponding complex conjugate coefficients, respectively. Here, 
$\sigma_a$ ($a = 1,2,3$) are the Pauli matrices. The selfdual 
and anti--selfdual generators of the $SO(4)$ rotations,
$(\sigma_{\mu\nu})_{AB}$ and $(\sigma_{\mu\nu})_{\dot{A}\dot{B}}$, obey
the relations
\begin{align}
\label{3.3}
&(\sigma_\mu)^{A \dot{C}} (\sigma_\nu)_{\dot{C}}^{\!~~B} = 
(\sigma_{\mu\nu})^{AB} - 
\delta_{\mu\nu} \epsilon^{AB},
\nonumber
\\
&(\sigma_\rho)^{A \dot{C}} (\sigma_{\mu\nu})_{\dot{C}}^{\!~~\dot{B}} =
\delta_{\rho\mu} (\sigma_\nu)^{A \dot{B}} -
\delta_{\rho\nu} (\sigma_\mu)^{A \dot{B}} -
\epsilon_{\mu\nu\rho\sigma} (\sigma^\sigma)^{A \dot{B}},
\\
\label{3.4}
&(\sigma_\mu)_{\dot{A} C} (\sigma_\nu)^C_{\!~~\dot{B}} = 
(\sigma_{\mu\nu})_{\dot{A}\dot{B}} + 
\delta_{\mu\nu} \epsilon_{\dot{A}\dot{B}},
\nonumber
\\
&(\sigma_\rho)_{\dot{A} C} (\sigma_{\mu\nu})^C_{\!~~B} =
\delta_{\rho\mu} (\sigma_\nu)_{\dot{A} B} -
\delta_{\rho\nu} (\sigma_\mu)_{\dot{A} B} +
\epsilon_{\mu\nu\rho\sigma} (\sigma^\sigma)_{\dot{A} B}.
\end{align}
The spinor index $A$ (and analogously $\dot{A}$) is raised and lowered as 
follows: $\epsilon^{AC} \varphi_C^{\!~~B} = \varphi^{AB}$ and 
$\varphi_A^{\!~~C} \epsilon_{CB} = \varphi_{AB}$,
where $\epsilon_{AB}$ (and analogous $\epsilon_{\dot{A}\dot{B}}$) is 
the invariant tensor of the group $SU(2)$, $\epsilon_{12} = \epsilon^{12} = 
\epsilon_{\dot{1}\dot{2}} = \epsilon^{\dot{1}\dot{2}} = 1$.

The action (\ref{3.1}) is manifestly invariant under hermitean
conjugation:
\begin{equation*}
( A_\mu, \lambda_{A \alpha}, \bar{\lambda}^{\dot{A} \alpha}, 
G^{\alpha\beta} ) \rightarrow
( - A_\mu, \bar{\lambda}_{\dot{A}}^{\!~~\alpha}, \lambda^A_{\!~~\alpha}, 
G_{\alpha\beta} ).
\end{equation*}
Furthermore, making use of (\ref{3.3}) and (\ref{3.4}), one verifies that 
(\ref{3.1}) is invariant also under the following on--shell supersymmetry 
transformations,
\begin{align*}
&Q_A^{\!~~\alpha} A_\mu = - i (\sigma_\mu)_{A \dot{B}} 
\bar{\lambda}^{\dot{B} \alpha}, 
\\
&Q_A^{\!~~\alpha} \bar{\lambda}_{\dot{B}}^{\!~~\beta} = 
(\sigma^\mu)_{A \dot{B}} D_\mu G^{\alpha\beta},
\\
&Q_A^{\!~~\alpha} G_{\beta\gamma} = 
2 i ( \delta^\alpha_{~\beta} \lambda_{A \gamma} - 
\delta^\alpha_{~\gamma} \lambda_{A \beta} ), 
\\
&Q_A^{\!~~\alpha} \lambda_{B \beta} = 
- \hbox{$\frac{1}{2}$} \delta^\alpha_{~\beta} 
(\sigma^{\mu\nu})_{AB} F_{\mu\nu} - 
\hbox{$\frac{1}{2}$} \epsilon_{AB} [ G^{\alpha\gamma}, G_{\gamma\beta} ] 
\end{align*}
and 
\begin{align*}
&\bar{Q}_{\dot{A} \alpha} A_\mu = i (\sigma_\mu)_{\dot{A} B} 
\lambda^B_{\!~~\alpha}, 
\\
&\bar{Q}_{\dot{A} \alpha} \lambda_{B \beta} = 
(\sigma^\mu)_{\dot{A} B} D_\mu G_{\alpha\beta},
\\
&\bar{Q}_{\dot{A} \alpha} G^{\beta\gamma} = 
2 i ( \delta_\alpha^{~\beta} \bar{\lambda}_{\dot{A}}^{\!~~\gamma} - 
\delta_\alpha^{~\gamma} \bar{\lambda}_{\dot{A}}^{\!~~\beta} ), 
\\
&\bar{Q}_{\dot{A} \alpha} \bar{\lambda}_{\dot{B}}^{\!~~\beta} = 
- \hbox{$\frac{1}{2}$} \delta_\alpha^{~\beta} 
(\sigma^{\mu\nu})_{\dot{A}\dot{B}} F_{\mu\nu} + 
\hbox{$\frac{1}{2}$} \epsilon_{\dot{A}\dot{B}} 
[ G_{\alpha\gamma}, G^{\gamma\beta} ]. 
\end{align*}
Let us recall that it is not possible to complete this superalgebra off--shell
with a finite number of auxiliary fields \cite{12}. 

In order to perform in (\ref{3.1}) the dimensional reduction to $D = 2$ we
re--name the third and fourth component of $A_\mu$ according to
\begin{equation}
\label{3.5}
A_3 = \hbox{$\frac{1}{2}$} ( \phi + \bar{\phi} ),
\qquad
A_4 = \hbox{$\frac{1}{2}$} i ( \phi - \bar{\phi} ),
\end{equation}
reserving the notation $A_\mu$ ($\mu = 1,2$) for the gauge field in $D = 2$.
Moreover, we decompose the components of $(\sigma_\mu)_{\dot{A}}^{\!~~B}$, 
$(\sigma_{\mu\nu})_{\dot{A}}^{\!~~\dot{B}}$ and 
$(\sigma_\mu)_A^{\!~~\dot{B}}$, $(\sigma_{\mu\nu})_A^{\!~~B}$ in the 
following manner,
\begin{alignat}{2}
\label{3.6}
&(\sigma_\mu)_{\dot{A}}^{\!~~B} 
\rightarrow 
i (\sigma_\mu)_A^{\!~~B},
&\qquad
&(\sigma_\mu)_A^{\!~~\dot{B}} 
\rightarrow 
i (\sigma_\mu)_A^{\!~~B},
\nonumber
\\
&(\sigma_3)_{\dot{A}}^{\!~~B} 
\rightarrow 
- i (\sigma_3)_A^{\!~~B},
&\qquad
&(\sigma_3)_A^{\!~~\dot{B}} 
\rightarrow 
- i (\sigma_3)_A^{\!~~B},
\nonumber
\\
&(\sigma_4)_{\dot{A}}^{\!~~B} 
\rightarrow 
\delta_A^{\!~~B},
&\qquad
&(\sigma_4)_A^{\!~~\dot{B}} 
\rightarrow 
- \delta_A^{\!~~B},
\nonumber
\\
&(\sigma_{\mu\nu})_{\dot{A}}^{\!~~\dot{B}}
\rightarrow 
- i \epsilon_{\mu\nu} (\sigma_3)_A^{\!~~B},
&\qquad
&(\sigma_{\mu\nu})_A^{\!~~B}
\rightarrow 
i \epsilon_{\mu\nu} (\sigma_3)_A^{\!~~B},
\nonumber
\\
&(\sigma_{\mu 3})_{\dot{A}}^{\!~~\dot{B}}
\rightarrow 
i \epsilon_{\mu\nu} (\sigma^\nu)_A^{\!~~B},
&\qquad
&(\sigma_{\mu 3})_A^{\!~~B}
\rightarrow 
- i \epsilon_{\mu\nu} (\sigma^\nu)_A^{\!~~B},
\nonumber
\\
&(\sigma_{\mu 4})_{\dot{A}}^{\!~~\dot{B}}
\rightarrow 
i (\sigma_\mu)_A^{\!~~B},
&\qquad
&(\sigma_{\mu 4})_A^{\!~~B}
\rightarrow 
i (\sigma_\mu)_A^{\!~~B},
\nonumber
\\
&(\sigma_{34})_{\dot{A}}^{\!~~\dot{B}}
\rightarrow 
- i (\sigma_3)_A^{\!~~B},
&\qquad
&(\sigma_{34})_A^{\!~~B}
\rightarrow 
- i (\sigma_3)_A^{\!~~B},
\end{alignat}
such that both the relations (\ref{3.3}) and (\ref{3.4}) become the
algebra of the Pauli matrices, 
\begin{align*}
&(\sigma_\mu)_A^{\!~~C} (\sigma_\nu)_{CB} = \delta_{\mu\nu} \epsilon_{AB} +
i \epsilon_{\mu\nu} (\sigma_3)_{AB},
\qquad
(\sigma_\mu, \sigma_3)_A^{\!~~B} = (\sigma_1, \sigma_2, \sigma_3),
\\
&(\sigma_\mu)_A^{\!~~C} (\sigma_3)_{CB} = - i \epsilon_{\mu\nu} 
(\sigma^\nu)_{AB},
\\
&(\sigma_3)_A^{\!~~C} (\sigma_3)_{CB} = \epsilon_{AB}.
\end{align*}

Then, from (\ref{3.1}) we obtain the Euclidean action of the $N = 16$, $D = 2$ SYM
\begin{align}
\label{3.7}
S^{(N = 16)} = \int_E d^2x\, {\rm tr}& \Bigr\{
\hbox{$\frac{1}{4}$} F_{\mu\nu} F^{\mu\nu} + 
\hbox{$\frac{1}{2}$} D_\mu \bar{\phi} D^\mu \phi -
\hbox{$\frac{1}{8}$} [ \bar{\phi}, \phi ]^2
\nonumber
\\
& - \hbox{$\frac{1}{2}$} \bar{\lambda}_A^{\!~~\alpha} (\sigma_3)^{AB}
[ \phi + \bar{\phi}, \lambda_{B \alpha} ] + 
\hbox{\large$\frac{1}{2}$} \bar{\lambda}^{A \alpha} 
[ \phi - \bar{\phi}, \lambda_{A \alpha} ]
\nonumber
\\
& + \bar{\lambda}_A^{\!~~\alpha} (\sigma_\mu)^{AB} 
D^\mu \lambda_{B \alpha} - 
\hbox{$\frac{1}{2}$} i \lambda_{A \alpha} 
[ G^{\alpha\beta}, \lambda^A_{\!~~\beta} ] -
\hbox{$\frac{1}{2}$} i \bar{\lambda}^{A \alpha} 
[ G_{\alpha\beta}, \bar{\lambda}_A^{\!~~\beta} ]
\phantom{\frac{1}{2}}
\nonumber
\\
& + \hbox{$\frac{1}{8}$} D_\mu G_{\alpha\beta}~D^\mu G^{\alpha\beta} + 
\hbox{$\frac{1}{8}$} [ \bar{\phi}, G_{\alpha\beta} ] [ \phi, G^{\alpha\beta} ] + 
\hbox{$\frac{1}{64}$} [ G_{\alpha\beta}, G_{\gamma\delta} ]
[ G^{\alpha\beta}, G^{\gamma\delta} ] \Bigr\}.
\end{align}
Since the decompositions (\ref{3.6}) explicitly include various factors of $i$, 
the action (\ref{3.7}) is no longer manifestly invariant under hermitean
conjugation. Rather, it is invariant under the following $Z_2$ symmetry,
\begin{equation}
\label{3.8}
Z_2:
\qquad
( A_\mu, \phi, \bar{\phi}, \lambda_{A \alpha}, \bar{\lambda}^{A \alpha}, 
G^{\alpha\beta} ) \rightarrow
( A_\mu, \bar{\phi}, \phi, - \bar{\lambda}_A^{\!~~\alpha}, - \lambda^A_{\!~~\alpha}, 
G_{\alpha\beta} ).
\end{equation}
Denoting the $N = 16$ spinorial supercharges in $D = 2$ by
$Q_A^{\!~~\alpha}$ and $\bar{Q}_{A \alpha}$, which are interchanged by the $Z_2$ 
symmetry (\ref{3.8}), the transformation rules of the re--named fields are:
\begin{align}
\label{3.9}
&Q_A^{\!~~\alpha} A_\mu = (\sigma_\mu)_{AB} \bar{\lambda}^{B \alpha}, 
\nonumber
\\
&Q_A^{\!~~\alpha} \phi = - (\sigma_3)_{AB} \bar{\lambda}^{B \alpha} -
\bar{\lambda}_A^{\!~~\alpha}, 
\nonumber
\\
&Q_A^{\!~~\alpha} \bar{\phi} = - (\sigma_3)_{AB} \bar{\lambda}^{B \alpha} +
\bar{\lambda}_A^{\!~~\alpha}, 
\nonumber
\\
&Q_A^{\!~~\alpha} \bar{\lambda}_B^{\!~~\beta} =
\hbox{$\frac{1}{2}$} i (\sigma^\mu)_{AB} D_\mu G^{\alpha\beta} - 
\hbox{$\frac{1}{2}$} i (\sigma_3)_{AB} [ \phi + \bar{\phi}, G^{\alpha\beta} ] - 
\hbox{$\frac{1}{2}$} i \epsilon_{AB} [ \phi - \bar{\phi}, G^{\alpha\beta} ],
\nonumber
\\
&Q_A^{\!~~\alpha} G_{\beta\gamma} = 
2 i ( \delta^\alpha_{~\beta} \lambda_{A \gamma} - 
\delta^\alpha_{~\gamma} \lambda_{A \beta} ), 
\nonumber
\\
&Q_A^{\!~~\alpha} \lambda_{B \beta} =  
\hbox{$\frac{1}{2}$} i \delta^\alpha_{~\beta}
\epsilon^{\mu\nu} (\sigma_\nu)_{AB} D_\mu ( \phi + \bar{\phi} ) +
\hbox{$\frac{1}{2}$} \delta^\alpha_{~\beta} 
(\sigma^\mu)_{AB} D_\mu ( \phi - \bar{\phi} )
\nonumber
\\
&\phantom{Q_A^{\!~~\alpha} \lambda_{B \beta} =} 
+ \hbox{\large$\frac{1}{2}$} \delta^\alpha_{~\beta} 
(\sigma^3)_{AB} [ \phi, \bar{\phi} ] - \hbox{$\frac{1}{2}$} i \delta^\alpha_{~\beta} 
\epsilon^{\mu\nu} (\sigma_3)_{AB} F_{\mu\nu} - 
\hbox{$\frac{1}{2}$} \epsilon_{AB} [ G^{\alpha\gamma}, G_{\gamma\beta} ].
\end{align}

\bigskip
%%%%%%%%%%%%%%%%%%%%%%%%%%%%%%%%%%%%%%%%%%%%%%%%%%%%%%%%%%%%%%%%%%%%%%%%%%%%%
\begin{flushleft}
{\large{\bf 4. $N_T=8$ topological twist of the $N=16$, $D=2$ super Yang--Mills
theory}}
\end{flushleft}
%%%%%%%%%%%%%%%%%%%%%%%%%%%%%%%%%%%%%%%%%%%%%%%%%%%%%%%%%%%%%%%%%%%%%%%%%%%%%
\medskip

Let us now perform the $N_T = 8$ topological twist of the $N = 16$, $D = 2$
SYM (for the group theoretical description of that topological twist 
we refer to \cite{3}). For that purpose we introduce the following set of 
twisted fields:
A $SU(4)$--quartet of Grassmann--odd vector fields $\psi_\mu^\alpha$, two 
$SU(4)$--quartets of Grassmann--odd scalar fields, $\bar{\eta}_\alpha$ and 
$\bar{\zeta}_\alpha$ which transform as the fundamental and its complex 
conjugate representation of $SU(4)$, respectively, and a $SU(4)$--sextet 
of Grassmann--even complex scalar fields $M_{\alpha\beta} = 
\hbox{$\frac{1}{2}$} \epsilon_{\alpha\beta\gamma\delta} M^{\gamma\delta}$, 
which transform as the second--rank complex selfdual representation of 
$SU(4)$.

Explicitly, the relationships between the original and the twisted fields
are chosen as follows:
\begin{align}
\label{3.12}
&\lambda_{A \alpha} = \hbox{$\frac{1}{2}$} \begin{pmatrix}
i (\sigma^\mu)_{AB} ( \psi_\mu^1 - \epsilon_{\mu\nu} \psi^{\nu 3} ) + 
(\sigma_3)_{AB} ( \bar{\eta}_4 + \bar{\zeta}_2 ) + 
i \epsilon_{AB} ( \bar{\zeta}_4 - \bar{\eta}_2 ) \\
i (\sigma^\mu)_{AB} ( \psi_\mu^2 + \epsilon_{\mu\nu} \psi^{\nu 4} ) + 
(\sigma_3)_{AB} ( \bar{\eta}_3 - \bar{\zeta}_1 ) + 
i \epsilon_{AB} ( \bar{\zeta}_3 + \bar{\eta}_1 ) \end{pmatrix},
\nonumber
\\
&\bar{\lambda}^{A \alpha} = \hbox{$\frac{1}{2}$} \begin{pmatrix}
i (\sigma^\mu)^{AB} ( \epsilon_{\mu\nu} \psi^{\nu 4} - \psi_\mu^2 ) + 
(\sigma_3)^{AB} ( \bar{\zeta}_1 + \bar{\eta}_3 ) -
i \epsilon^{AB} ( \bar{\eta}_1 - \bar{\zeta}_3 ) \\
i (\sigma^\mu)^{AB} ( \epsilon_{\mu\nu} \psi^{\nu 3} + \psi_\mu^1 ) + 
(\sigma_3)^{AB} ( \bar{\zeta}_2 - \bar{\eta}_4 ) - 
i \epsilon^{AB} ( \bar{\eta}_2 + \bar{\zeta}_4 ) \end{pmatrix},
\\
\intertext{between $\lambda_{A \alpha}$, $\bar{\lambda}^{A \alpha}$ and the 
twisted vector and scalar fields $\psi_\mu^\alpha$, $\bar{\eta}_\alpha$, 
$\bar{\zeta}_\alpha$, as well as}
\label{3.13}
&\phi = M_1 - i M_2,
\nonumber
\\
&\bar{\phi} = M_1 + i M_2,
\\
\label{3.14}
&G_{\alpha\beta} = \begin{pmatrix}
\epsilon_{AB} M_6 & - (\sigma^\mu)_{AB} V_\mu +
(\sigma_3)_{AB} M_3 - i \epsilon_{AB} M_4 \\
(\sigma^\mu)_{AB} V_\mu - (\sigma_3)_{AB} M_3 - 
i \epsilon_{AB} M_4 & \epsilon_{AB} M_5
\end{pmatrix},
\nonumber
\\
&G^{\alpha\beta} = \begin{pmatrix}
\epsilon^{AB} M_5 & (\sigma^\mu)^{AB} V_\mu - 
(\sigma_3)^{AB} M_3 + i \epsilon^{AB} M_4 \\
- (\sigma^\mu)^{AB} V_\mu + (\sigma_3)^{AB} M_3 + 
i \epsilon^{AB} M_4 & \epsilon^{AB} M_6 
\end{pmatrix},
\\
\intertext{where}
&M_1 = \hbox{$\frac{1}{2}$} ( M^{12} + M^{34} ),
\qquad
M_3 = \hbox{$\frac{1}{2}$} i ( M^{12} - M^{34} ),
\qquad
M_5 = M^{24},
\nonumber
\\
&M_2 = \hbox{$\frac{1}{2}$} ( M^{14} + M^{23} ),
\qquad
M_4 = \hbox{$\frac{1}{2}$} i ( M^{14} - M^{23} ),
\qquad
M_6 = M^{31},
\nonumber
\end{align}
between $\phi$, $\bar{\phi}$, $G_{\alpha\beta}$ and the twisted vector and 
scalar fields $V_\mu$ and $M_{\alpha\beta}$, respectively.

Thereby, the assignment between the index $\alpha$ of the internal group 
 and the spinor index $A$ is the following: In (\ref{3.12}) the spinor 
indices $B = 1,2$ at the top and at the bottom of both columns correspond 
to the values $\alpha = 1,2$ and $\alpha = 3,4$ of both spinors 
$\lambda_{A \alpha}$ and $\bar{\lambda}^{A \alpha}$, respectively. Similary, 
in (\ref{3.14}) the spinor indices $A = 1,2$ (resp. $B = 1,2$) at the upper 
and at the lower raw (resp. at the left and at the right column) of the both 
matrices correspond to the values $\alpha = 1,2$ and $\alpha = 3,4$ 
(resp. $\beta = 1,2$ and $\beta = 3,4$) of the scalar fields 
$G_{\alpha\beta}$, respectiely. By using the explicit form (\ref{3.2}) of the 
Clebsch--Gordon coefficients one establishes that $G_{\alpha\beta}$ and 
$G^{\alpha\beta}$ in (\ref{3.14}) are actually dual to each other, 
$G_{\alpha\beta} = \hbox{$\frac{1}{2}$} 
\epsilon_{\alpha\beta\gamma\delta} G^{\gamma\delta}$. 

The relationship between the spinorial supercharges $Q^{A \alpha}$ and
$\bar{Q}_{A \alpha}$, being interrelated by the $Z_2$ symmetry (\ref{3.8}), 
and the twisted scalar and vector supercharges $Q^\alpha$, $^\star Q^\alpha$ 
and $\bar{Q}_{\mu \alpha}$, being interchanged by a discrete Hodge--type 
$\star$ operation (see Eq. (\ref{3.17}) below), is quite similar to the ones 
of the spinor fields, Eq. (\ref{3.12}), namely
\begin{align}
\label{3.15}
&Q^{A \alpha} = \hbox{$\frac{1}{2}$} \begin{pmatrix}
i (\sigma^\mu)^{AB} ( \bar{Q}_{\mu 1} - \epsilon_{\mu\nu} \bar{Q}^\nu_3 ) - 
(\sigma_3)^{AB} ( Q^4 - i \,^\star Q^2 ) - 
\epsilon^{AB} ( \,^\star Q^4 - i Q^2 ) \\
i (\sigma^\mu)^{AB} ( \bar{Q}_{\mu 2} + \epsilon_{\mu\nu} \bar{Q}^\nu_4 ) - 
(\sigma_3)^{AB} ( Q^3 + i \,^\star Q^1 ) - 
\epsilon^{AB} ( \,^\star Q^3 + i Q^1) \end{pmatrix},
\nonumber
\\
&\bar{Q}_{A \alpha} = \hbox{$\frac{1}{2}$} \begin{pmatrix}
i (\sigma^\mu)_{AB} ( \epsilon_{\mu\nu} \bar{Q}^\nu_4 - \bar{Q}_{\mu 2} ) + 
(\sigma_3)_{AB} ( i \,^\star Q^1 - Q^3 ) +
\epsilon_{AB} ( i Q^1 - \,^\star Q^3 ) \\
i (\sigma^\mu)_{AB} ( \epsilon_{\mu\nu} \bar{Q}^\nu_3 + \bar{Q}_{\mu 1} ) + 
(\sigma_3)_{AB} ( i \,^\star Q^2 + Q^4 ) + 
\epsilon_{AB} ( i Q^2 + \,^\star Q^4 ) \end{pmatrix}.
\end{align}

After performing in (\ref{3.7}) the topological twist 
(\ref{3.12}) -- (\ref{3.14}) and introducing the Grassmann--even
auxiliary fields $B$, $\bar{B}$, $Y$ and $E_{\mu \alpha\beta} =
\frac{1}{2} \epsilon_{\alpha\beta\gamma\delta} E_\mu^{\gamma\delta}$ one gets 
the following $N_T = 8$ Hodge--type cohomological gauge theory with global 
symmetry group $SU(4)$:
\begin{align}
\label{3.16}
S^{(N_T = 8)} = \int_E d^2x\, {\rm tr}& \Bigr\{ 
\hbox{$\frac{1}{4}$} i \epsilon^{\mu\nu}
B F_{\mu\nu}(A + i V) - \hbox{$\frac{1}{4}$} i \epsilon^{\mu\nu} 
\bar{B} F_{\mu\nu}(A - i V) - \hbox{$\frac{1}{2}$} \bar{B} B
\nonumber
\\
& - \epsilon^{\mu\nu} \bar{\zeta}_\alpha D_\mu(A + i V) \psi_\nu^\alpha - 
\bar{\eta}_\alpha D^\mu(A - i V) \psi_\mu^\alpha -
\hbox{$\frac{1}{4}$} E^\mu_{\alpha\beta} E_\mu^{\alpha\beta}
\nonumber
\\
& + \hbox{$\frac{1}{2}$} i \epsilon^{\mu\nu} 
M_{\alpha\beta} \{ \psi_\mu^\alpha, \psi_\nu^\beta \} + 
i M^{\alpha\beta} \{ \bar{\eta}_\alpha, \bar{\zeta}_\beta \} -
Y D^\mu(A) V_\mu - \hbox{$\frac{1}{2}$} Y^2
\phantom{\frac{1}{2}}
\nonumber
\\
& + \hbox{$\frac{1}{8}$}
D^\mu(A + i V) M_{\alpha\beta}~D_\mu(A - i V) M^{\alpha\beta} +
\hbox{$\frac{1}{64}$} [ M_{\alpha\beta}, M_{\gamma\delta} ] 
[ M^{\alpha\beta}, M^{\gamma\delta} ] \Bigr\}.
\end{align}
In this $SU(4)$ symmetric form the action (\ref{3.16}) is manifestly
inariant under the following Hodge--type $\star$ symmetry, 
defined by the replacements
\begin{equation}
\label{3.17}
\varphi \equiv \begin{bmatrix}
\partial_\mu & A_\mu & V_\mu &  
\\
\psi_\mu^\alpha & \bar{\eta}_\alpha & \bar{\zeta}_\alpha & M^{\alpha\beta}  
\\
B & \bar{B} & Y & E_\mu^{\alpha\beta} 
\end{bmatrix}
\quad \Rightarrow \quad
\star \varphi = \begin{bmatrix} 
\epsilon_{\mu\nu} \partial^\nu & \epsilon_{\mu\nu} A^\nu & 
- \epsilon_{\mu\nu} V^\nu &  
\\
- i \psi_\mu^\alpha & - i \bar{\zeta}_\alpha & i \bar{\eta}_\alpha & 
- M^{\alpha\beta} 
\\
- \bar{B} & - B & - Y & \epsilon_{\mu\nu} E^{\nu \alpha\beta} 
\end{bmatrix}.
\end{equation}
with the property $\star ( \star \varphi ) = - P \varphi$. Here, $P$ is the
operator of Grassmann--parity whose eigenalues are defined by
\begin{equation*}
P \varphi = \begin{cases} + \varphi & \text{if $\varphi$ is Grassmann-odd},
            \\            - \varphi & \text{if $\varphi$ is Grassmann-even}
            \end{cases}~.
\end{equation*}
Hence, after twisting the $Z_2$ symmetry (\ref{3.8}) changes into the
Hodge--type $\star$ symmetry (\ref{3.17}).

The transformations rules for the topological shift symmetry,
generated by $Q^\alpha$, are
\begin{align}
\label{3.18}
&Q^\alpha A_\mu = \psi_\mu^\alpha,
\nonumber
\\
&Q^\alpha V_\mu = - i \psi_\mu^\alpha,
\nonumber
\\
&Q^\alpha M_{\beta\gamma} = 2 i (
\delta^\alpha_{~\beta} \bar{\zeta}_\gamma -
\delta^\alpha_{~\gamma} \bar{\zeta}_\beta ),
\nonumber
\\
&Q^\alpha \psi_\mu^\beta = E_\mu^{\alpha\beta} -
i \epsilon_{\mu\nu} D^\nu(A - i V) M^{\alpha\beta},
\nonumber
\\
&Q^\alpha \bar{\zeta}_\beta = i \delta^\alpha_{~\beta} B,
\nonumber
\\
&Q^\alpha B = 0,
\nonumber
\\
&Q^\alpha \bar{\eta}_\beta = i \delta^\alpha_{~\beta} Y +
\hbox{$\frac{1}{2}$} [ M^{\alpha\gamma}, M_{\gamma\beta} ],
\nonumber
\\
&Q^\alpha Y = [ M^{\alpha\beta}, \bar{\zeta}_\beta ],
\nonumber
\\
&Q^\alpha \bar{B} = - 2 [ M^{\alpha\beta}, \bar{\eta}_\beta ],
\nonumber
\\
&Q^\alpha E^\mu_{\beta\gamma} = \delta^\alpha_{~[\beta} \bigr(
\epsilon^{\mu\nu} D_\nu(A + i V) \bar{\zeta}_{\gamma]} - 
D^\mu(A - i V) \bar{\eta}_{\gamma]} - 
i \epsilon^{\mu\nu} [ M_{\gamma]\delta}, \psi_\nu^\delta ] \bigr). 
\end{align}
From combining $Q^\alpha$ with the above displayed Hodge--type $\star$
symmetry one gets the corresponding transformations rules for the topological 
co--shift symmetry: $^\star Q^\alpha = P \star Q^\alpha \star$. 

By a straightforward calculation one verifies that  both the
supercharges $Q^\alpha$ and $^\star Q^\alpha$ provide an {\it off--shell} 
realization of the following topological superalgebra,
\begin{eqnarray}
\label{X}
\{ Q^\alpha, Q^\beta \} = 0,
\qquad
\{ Q^\alpha, \,^\star Q^\beta \} = - 2 \delta_G(M^{\alpha\beta}),
\qquad
\{ \,^\star Q^\alpha, \,^\star Q^\beta \} = 0,
\end{eqnarray}
where the field--dependent gauge transformations are defined by 
$\delta_G(M^{\alpha\beta}) A_\alpha = - D_\alpha M^{\alpha\beta}$ and 
$\delta_G(M^{\alpha\beta}) X = [ M^{\alpha\beta}, X ]$ for all the
other fields. 

Obviously, the structure of this superalgebra is directly analogous to the 
de Rham cohomology in differential geometry: The exterior and the
co--exterior derivatives $d$ and $\delta = \pm \star d \star$, being 
interrelated by the duality $\star$ operation, correspond to the nilpotent 
topological shift and co--shift operators $Q^\alpha$ and $^\star Q^\alpha =
P \star Q^\alpha \star$, respectively. Moreover, the Laplacian $\Delta = 
\{ d, \delta \}$ corresponds to the field--dependent gauge generator 
$\delta_G(M^{\alpha\beta})$, so that we have indeed a perfect example of a 
Hodge--type cohomological gauge theory. 

Furthermore, by an explicit calculation one can verify that the action (\ref{3.16}) is
also invariant under the following {\it on--shell} vector supersymmetries,
\begin{align}
\label{3.19}
&\bar{Q}_{\mu \alpha} A_\nu = \delta_{\mu\nu} \bar{\eta}_\alpha -
\epsilon_{\mu\nu} \bar{\zeta}_\alpha,
\nonumber
\\
&\bar{Q}_{\mu \alpha} V_\nu = - i \delta_{\mu\nu} \bar{\eta}_\alpha -
i \epsilon_{\mu\nu} \bar{\zeta}_\alpha,
\nonumber
\\
&\bar{Q}_{\mu \alpha} M^{\beta\gamma} = 2 i \epsilon_{\mu\nu} (
\delta_\alpha^{~\beta} \psi^{\nu \gamma} -
\delta_\alpha^{~\gamma} \psi^{\nu \beta} ),
\nonumber
\\
&\bar{Q}_{\mu \alpha} \bar{\zeta}_\beta = \epsilon_{\mu\nu} 
E^\nu_{\alpha\beta} + i D_\mu(A - i V) M_{\alpha\beta},
\nonumber
\\
&\bar{Q}_{\mu \alpha} \bar{\eta}_\beta = E_{\mu \alpha\beta} + 
i \epsilon_{\mu\nu} D^\nu(A + i V) M_{\alpha\beta},
\nonumber
\\
&\bar{Q}_{\mu \alpha} \psi_\nu^\beta = - 2 \delta_\alpha^{~\beta} F_{\mu\nu}(A) - 
2 i \delta_\alpha^{~\beta} D_\mu(A) V_\nu -
i \delta_\alpha^{~\beta} \delta_{\mu\nu} Y - 
i \delta_\alpha^{~\beta} \epsilon_{\mu\nu} \bar{B} +
\hbox{$\frac{1}{2}$} \delta_{\mu\nu} [ M_{\alpha\gamma}, M^{\gamma\beta} ],
\nonumber
\\
&\bar{Q}_{\mu \alpha} \bar{B} = 2 i \epsilon_{\mu\nu}
D^\nu(A + i V) \bar{\eta}_\alpha,
\nonumber
\\
&\bar{Q}_{\mu \alpha} Y = 2 i D_\mu(A - i V) \bar{\eta}_\alpha -
\epsilon_{\mu\nu} [ M_{\alpha\beta}, \psi^{\nu \beta} ],
\nonumber
\\
&\bar{Q}_{\mu \alpha} B = 2 i \epsilon_{\mu\nu} D^\nu(A - i V) \bar{\eta}_\alpha + 
4 i D_\mu(A) \bar{\zeta}_\alpha + 2 [ M_{\alpha\beta}, \psi_\mu^\beta ],
\nonumber
\\
&\bar{Q}_{\mu \alpha} E_\nu^{\beta\gamma} = - \delta_\alpha^{~[\beta}
D_{[\mu}(A + i V) \psi_{\nu]}^{~\gamma]} - 
\delta_\alpha^{~[\beta} \delta_{\mu\nu} D^\rho(A - i V) \psi_\rho^{\gamma]} +
i \delta_\alpha^{~[\beta} [ M^{\gamma] \delta}, 
\epsilon_{\mu\nu} \bar{\eta}_\delta - 
\delta_{\mu\nu} \bar{\zeta}_\delta ].
\end{align}
In addition, this action is also invariant under the co--vector supersymmetries
\begin{equation*}
^\star \bar{Q}_{\mu \alpha} = P \star \bar{Q}_{\mu \alpha} \star \doteq 
i \bar{Q}_{\mu \alpha},
\end{equation*}
which on--shell, i.e., by using {\it only} the equations of motion of the
auxiliarly fields, become $i$ times the vector supersymmetries! Hence, it
holds
\begin{equation*}
( Q^\alpha, \,^\star Q^\alpha, \bar{Q}_{\mu \alpha} ) S^{(N_T = 8)} = 0,
\end{equation*}
and the total number of (real) supercharges is actually $N = 16$. 

Finally, let us mention that there is also a $N_T = 4$ topological twist of 
$N = 16$, $D = 2$ SYM with global symmetry group $SO(4) \otimes SU(2)$. 
This topological theory can be regarded as the $N_T = 4$ super--BF theory
coupled to a spinorial hypermultiplet. Another way of obtaining the action
of this theory is to dimensionally reduce either the higher dimensional 
analogue of the Donaldson--Witten theory in $D = 8$ \cite{13,14} to $D = 2$ or
to dimensionally reduce the $N_T = 1$ half--twisted theory \cite{15} in 
$D = 4$ to $D = 2$. However, that topological twist does not lead to another
Hodge--type cohomological theory, since the underlying cohomology is only
equivariantly nilpotent and not strictly nilpotent as in Eqs.~(\ref{X}).

\end{document}